%Paper: hep-ph/9406392
%From: manohar@sphal.UCSD.EDU (Aneesh V. Manohar)
%Date: Fri, 24 Jun 1994 12:39:01 -0700

%%%%%%%%%%%%%%%%%%%%%%%%%%%%%%%%%%%%%%%%%%%%%%%%%%%%%%%%%%%%%%%%%%%
%                       INSTRUCTIONS
%
% This paper uses the harvmac macros. 2 files with postscript figures
% have been included as a uuencoded tar file with instructions for
% unpacking. If you have  epsf.tex, uncomment the following line
% and the postscript figures
%
%\input epsf
%
% will be included in the paper by the dvips program. If you do not
% have epsf.tex, you can print the figures out separately.
%
%%%%%%%%%%%%%%%%%%%%%%%%%%%%%%%%%%%%%%%%%%%%%%%%%%%%%%%%%%%%%%%%%%%%%%%%
\ifx\epsffile\undefined\message{(FIGURES WILL BE IGNORED)}
\def\insertfig#1{}% null macro
\else\message{(FIGURES WILL BE INCLUDED)}
\def\insertfig#1{{{
\midinsert\centerline{\epsfxsize=\hsize
\epsffile{#1}}\bigskip\bigskip\bigskip\bigskip\endinsert}}}
\fi

\input harvmac
%%%%%%%%%%%%%%%%%%%%%%%%%%%%%%%%%%%%%%%%%%%%%%%%%%%%%%%%%%%%%%%%%%%%%%
%
%  UCSD macros to overwrite some of the definitions in harvmac.tex
%  (include after harvmac.tex)
%  last modified 4/92
%
%%%%%%%%%%%%%%%%%%%%%%%%%%%%%%%%%%%%%%%%%%%%%%%%%%%%%%%%%%%%%%%%%%%%%%%
%
% modify the output routine for the little format
%
\ifx\answ\bigans
\else
\output={
  \almostshipout{\leftline{\vbox{\pagebody\makefootline}}}\advancepageno
}
\fi
%
%
% address
%

%
% grant numbers
%

%
% preprint number
%
\def\UCSD#1#2{\noindent#1\hfill #2%
\bigskip\supereject\global\hsize=\hsbody%
\footline={\hss\tenrm\folio\hss}}% restores pagenumbers
%
% abstract
%
\def\abstract#1{\centerline{\bf Abstract}\nobreak\medskip\nobreak\par #1}
%
%
% titlefont
%
%
\edef\tfontsize{ scaled\magstep3}
 \tfontsize  \tfontsize
 \tfontsize \font\titlei=cmmi10 \tfontsize
\font\titleis=cmmi7 \tfontsize \font\titleiss=cmmi5 \tfontsize
\font\titlesy=cmsy10 \tfontsize \font\titlesys=cmsy7 \tfontsize
\font\titlesyss=cmsy5 \tfontsize  \tfontsize
\skewchar\titlei='177 \skewchar\titleis='177 \skewchar\titleiss='177
\skewchar\titlesy='60 \skewchar\titlesys='60 \skewchar\titlesyss='60
%
%\def\titlefont{\def\rm{\fam0\titlerm}% switch to title font
%\textfont0=\titlerm \scriptfont0=\titlerms \scriptscriptfont0=\titlermss
%\textfont1=\titlei \scriptfont1=\titleis \scriptscriptfont1=\titleiss
%\textfont2=\titlesy \scriptfont2=\titlesys \scriptscriptfont2=\titlesyss
%\textfont\itfam=\titleit \def\it{\fam\itfam\titleit}\rm}
%
%
% math symbols
%
%---------------------------------------------------------------------
%
\def\inv{^{\raise.15ex\hbox{${\scriptscriptstyle -}$}\kern-.05em 1}}
  %prime
\def\lbar{{\lower.35ex\hbox{$\mathchar'26$}\mkern-10mu\lambda}} %lambda bar

%
%
% various slashed symbols
%
%
 % slashes a character
\def\dsl{\,\raise.15ex\hbox{/}\mkern-13.5mu D} %this one can be subscripted
\def\delsl{\raise.15ex\hbox{/}\kern-.57em\partial}
\def\Ksl{\hbox{/\kern-.6000em\rm K}}
\def\Asl{\hbox{/\kern-.6500em \rm A}}
\def\Dsl{\hbox{/\kern-.6000em\rm D}} %roman D
\def\Qsl{\hbox{/\kern-.6000em\rm Q}}
\def\gradsl{\hbox{/\kern-.6500em$\nabla$}}
%
% space and backspace in l mode
%
\def\lspace{\ifx\answ\bigans{}\else\qquad\fi}
\def\lbspace{\ifx\answ\bigans{}\else\hskip-.2in\fi} % $$\lbspace...$$
%
%     boxes an equation
%
\def\boxeqn#1{\vcenter{\vbox{\hrule\hbox{\vrule\kern3pt\vbox{\kern3pt
        \hbox{${\displaystyle #1}$}\kern3pt}\kern3pt\vrule}\hrule}}}
%
%     draw a little box (end of proof symbol)
%     e.g. \mbox{.1}{.1}
%
\def\mbox#1#2{\vcenter{\hrule \hbox{\vrule height#2in
\kern#1in \vrule} \hrule}}
%
%
%
%     curly letters
%
   %curly letters

  \def\CO{{\cal O}}

%
%
%
%     derivatives
%
%

%

\def\bra#1{\left\langle #1\right|}
\def\ket#1{\left| #1\right\rangle}
\def\abs#1{\left| #1\right|}

\def\darr#1{\raise1.5ex\hbox{$\leftrightarrow$}\mkern-16.5mu #1}

%
 %pound sterling
%
 %puts a small half in a displayed eqn
\def\frac#1#2{{\textstyle{#1\over #2}}} %puts a small fraction
%in a displayed eqn
%
%
%     various math operators
%
%
\def\tr{\mathop{\rm tr}}

%
%
%
%

%
%       relations
%
\def\ltap{\ \raise.3ex\hbox{$<$\kern-.75em\lower1ex\hbox{$\sim$}}\ }
\def\gtap{\ \raise.3ex\hbox{$>$\kern-.75em\lower1ex\hbox{$\sim$}}\ }
\def\gl{\ \raise.5ex\hbox{$>$}\kern-.8em\lower.5ex\hbox{$<$}\ }
\def\roughly#1{\raise.3ex\hbox{$#1$\kern-.75em\lower1ex\hbox{$\sim$}}}
%
%
%       This defines et al., i.e., e.g., cf., etc.

%

%
\def\np#1#2#3{Nucl. Phys. B{#1} (#2) #3}

\def\prl#1#2#3{Phys. Rev. Lett. {#1} (#2) #3}
\def\physrev#1#2#3{Phys. Rev. {#1} (#2) #3}

\relax

\noblackbox

\def\tr{{\rm Tr}}
\def\lqcd{\Lambda_{\rm QCD}}

\centerline{{\titlefont{Power Suppressed Corrections to}}}
\medskip
\centerline{{\titlefont{Hadronic Event Shapes}}}
\bigskip
\centerline{Aneesh V.~Manohar}
\smallskip
\centerline{{\it Department of Physics, University of California at San
Diego, La Jolla, CA 92093}}
\bigskip
\centerline{Mark B. Wise}
\smallskip
\centerline{{\it California Institute of Technology, Pasadena, CA  91125}}
\vfill
\abstract{For high energy processes ($M\gg \Lambda_{QCD}$) there are
infrared safe hadronic shape variables that have a calculable perturbative
expansion in $\alpha_s(M^2)$. However, nonperturbative power suppressed
corrections to these variables are not well understood. We use the behavior
of large orders of the perturbation expansion to gain insight into the
nonperturbative corrections. Our results suggest that certain shape variables
have nonperturbative corrections suppressed by fractional powers of
$\Lambda_{QCD}^2/M^2$.
}
\vfill
%\draftmode
\UCSD{\vbox{\hbox{UCSD/PTH 94--11}\hbox{CALT--68--1937}
\hbox{hep-ph/9406392}}}{June 1994}

Many large momentum transfer processes involving the strong interactions have a
power series expansion in the strong coupling constant $\alpha_s (M^2)$. In
addition to this perturbative expansion in $\alpha_s$, there are
nonperturbative corrections suppressed by powers of $\lqcd^2/M^2$, where $M$ is
the typical momentum transfer, $M\gg \lqcd$. In some cases the nonperturbative
corrections are well understood.  For example, in $R(M^2) = \sigma (e^+ e^-
\rightarrow$ hadrons)$/\sigma (e^+ e^- \rightarrow \mu^+ \mu^-)$, with $M^2 =
(p_{e^{-}} + p_{e^{+}})^2$, the operator product expansion can be used to show
that nonperturbative power corrections are proportional to the vacuum
expectation values of gauge invariant local operators.\foot{There may
also be non-perturbative instanton contributions to the coefficients of
operators \ref\SVZ{M.A.~Shifman,
A.I.~Vainshtein, V.I.~Zakharov, \np {147} {1979} 385 }} The leading
nonperturbative correction is of order $\lqcd^4/M^4$, and arises from the
expectation value of the dimension four operator $\tr\, G_{\mu\nu} G^{\mu\nu}$,
 where $G_{\mu\nu}$ is the gluon field strength tensor. The operator product
expansion shows that there are no corrections of order $\lqcd^2/M^2$, since
there is no gauge invariant dimension two operator that can be constructed out
of quark and gluon fields. However, there are physical quantities that can be
predicted using perturbative QCD for which the nonperturbative corrections
are not well understood. For example, infrared safe event shape variables in
$e^+e^- \rightarrow$ hadrons have a perturbative expansion in $\alpha_s(M^2)$.
Since these variables weight the final hadronic states in a way that
depends on their shape, they are not given by the imaginary part of the time
ordered product of electromagnetic currents and an operator product expansion
does not seem feasible. In this letter we examine some
features of the nonperturbative corrections to hadronic shape variables
in a toy model.

The strong interaction coupling constant $\alpha_s(M^2)$ falls logarithmically
with the momentum transfer $M^2$. Nevertheless, perturbative QCD corrections
can also produce power suppressed corrections \ref\zakharov{V.I.~Zakharov,
\np {385} {1992} 452 } to physical quantities, because
the QCD perturbative expansion is an asymptotic series. Information on the
power suppressed corrections to physical quantities can be obtained from an
examination of the behavior of the perturbative expansion in $\alpha_s$ at
large orders.  Consider a physical quantity $P(\alpha_s)$ that has a power
series expansion
\eqn\pseries{
P(\alpha_s) = \sum_{n=0}^\infty p_n \alpha_s^{n + 1}
}
in the strong coupling $\alpha_s(M^2)$.  It's Borel transform is defined by
\eqn\pborel{
B[P](t) = \sum_{n=0}^\infty {p_n t^n\over n!}.
}
The Borel transform $B[P](t)$ may converge, even if the original series
$P(\alpha_s)$ is only an asymptotic expansion. This provides a definition of $P
(\alpha_s)$ through the integral transform
\eqn\binverse{
P (\alpha_s) = \int_0^\infty dt e^{-(t/\alpha_{s})} B[P] (t).
}
However, if the coefficients $p_n$ grow too fast with $n$, the Borel transform
$B[P]$ will have singularities that prevent using \binverse\ to define
$P(\alpha_s)$.  In an asymptotically free theory such as QCD, singularities in
$B[P] (t)$ for positive $t$ are associated with the infrared properties of
Feynman diagrams, and are usually referred to as infrared renormalons. One can
still obtain $P(\alpha_s)$ from $B[P](t)$ using the inverse Borel transform
eq.~\binverse, provided one deforms the contour of integration in
eq.~\binverse\ away from the real $t$ axis, to avoid the renormalon
singularity. The resultant expression for $P(\alpha_s)$ is no longer
unique (or even real),
and depends on whether the deformed path passes above or below the renormalon
pole. Suppose, for example, that $B[P] (t)$ has a simple pole at
\eqn\tpole{
t = {u\over b_0},
}
where $b_0$ is the first term in the $\beta$-function (negative in QCD),
\eqn\betafn{
\mu^2 {d\alpha_s\over d\mu^2} = b_0 \alpha_s^2 + \CO\left(\alpha_s^3\right),
}
which governs the large $M^2$ behavior of the coupling constant in an
asymptotically free theory,
\eqn\aasymp{
\alpha_s (M^2) \sim {1\over \left(-b_0\right) \ln\left(M^2/\lqcd^2\right)} ,
}
and $u<0$ is a constant (which is usually, but not always, an integer).
The magnitude of the ambiguity in eq.~\binverse\ associated with this
singularity can be estimated by the difference between the value of $P$
obtained by deforming the $t$ integration contour above and below the
pole \ref\zj{J.~Zinn-Justin, Phys. Rep. 49 (1979) 205}. The difference is
\eqn\pambig{\eqalign{
\abs{\Delta P} &\sim  e^{-u/b_{0}\alpha_{s}}\cr
&\sim \left(\lqcd^2/M^2\right)^{(-u)}.
}}
Thus infrared renormalons produce power
suppressed ambiguities in $P$. The infrared renormalon closest to the origin
$u=t=0$ gives the dominant ambiguity. For $R(M^2)$, the total cross-section for
$e^+e^-\rightarrow {\rm hadrons}$ defined previously, it is believed that the
perturbative expansion is not Borel summable, and has a renormalon at $u=-2$,
which gives rise to an ambiguity in $R$ that is of order $\lqcd^4/M^4$
\ref\gparisi{G.~Parisi, Phys. Lett. B76 (1978) 65, Nucl. Phys. B150
(1979) 163}\ref\amueller{A.H.~Mueller, Nucl. Phys. B250 (1985) 327}. This
ambiguity is of the same form as the leading non-perturbative correction in the
operator product expansion for $R(M^2)$, due to the vacuum expectation value of
$\tr\ G_{\mu\nu} G^{\mu\nu}$. It is thought that there is an ambiguity in the
definition of $\bra0 \tr\ G_{\mu\nu} G^{\mu\nu} \ket 0$, and that the
renormalon ambiguity in the sum of the
perturbation series cancels the ambiguity in the non-perturbative matrix
element, to give a well-defined result for the total cross-section
$R(M^2)$\ref\fdavid{F.~David, \np {209} {1982} {433}, \np {234} {1984}
{237}, \np {263} {1986} {637} }.

Renormalon ambiguities can only be absorbed into non-perturbative matrix
elements if the contributions of both quantities to a physical observable such
as $R(M^2)$ have the same $M$ dependence. Renormalon ambiguities can be
absorbed in gauge invariant matrix elements provided the operator dimension is
equal to $-2u$.  Explicit computations on renormalons are at a primitive level.
Typically, one sums bubble graphs like those given in \fig\bubble{Graphs
that give an infrared renormalon in QCD.} to determine the renormalon
singularities \ref\thooft{G.~'t Hooft, in {\it The Whys of Subnuclear Physics},
 edited by A.~Zichichi (Plenum, New York, 1978) }. Other graphs which have been
neglected are presumably just as important, so the result of summing the bubble
chain is not a proof of the existence of renormalon effects. Renormalon effects
have been computed by Beneke \ref\beneke{M.~Beneke, \np {405} {1993}
{424} }\ in a limiting case of QCD. One considers
QCD with $N_f$ flavors, and takes the limit $N_f\rightarrow\infty$ with
$a=N_f\alpha_s$ held fixed. Feynman diagrams are computed to leading order in
$\alpha_s$, but to all orders in $a$. Terms in the bubble sum of \bubble\ with
any number of bubbles are kept in this limit, since each additional
fermion loop contributes a factor $\alpha_s N_f$, which is not treated
 as small. The $N_f \rightarrow \infty$ limit has its limitations,
 since QCD is not an
asymptotically free theory  in this limit. The procedure used by Beneke
is to write the Borel transform as a function of $u=b_0 t$, (where $b_0$
is now positive), but still study renormalons for negative $u$. The
singularities at negative $u$ are then taken to be the infrared
renormalons for
asymptotically free QCD. This procedure was used by Beneke to study
renormalons in $R(M^2)$. It has been suggested by Brown and Yaffe
\ref\brownyaffe{L.S.~Brown and L.G.~Yaffe, \physrev {D45} {1992} {R398} \semi
L.S.~Brown, L.G.~Yaffe and C.~Zhai, \physrev {D46} {1992} {4712} }\ that
there might be a renormalon at $u=-1$, which would
lead to an order  $\lqcd^2/M^2$ ambiguity in the perturbative expansion
for $R(M^2)$ that could not be compensated by the matrix element of a
local operator. Beneke found by explicit computation that the renormalon
singularity at $u=-1$ vanished.

In this letter we examine the behavior of shape variables at large orders in
perturbation theory using a toy model.  The model consists of a $U_1 (1) \times
U_2(1)$ gauge group (with gauge couplings $e_1$ and $e_2$), and a large number
$N_f$ of massless fermions $f$ with charge (1,0). We suppose there is a neutral
scalar $\phi$ of mass $M$ that decays to the massless degrees of freedom
through the dimension five interaction Lagrangian density.
\eqn\efflag{
{\cal L}_{\rm int} = \lambda e_1 e_2\ \phi\ F_{\mu\nu}^{(1)}\ F^{(2) \mu\nu}
}
where $\lambda$ has mass dimension $-1$. We will compute the energy-energy
correlation \ref\ellis{C.L.~Basham, L.S.~Brown, S.D.~Ellis, and
S.T.~Love, \prl {41} {1978} {1585} }\ for $\phi$ decay to leading order in
$\alpha_1$ and $\alpha_2$, and to all orders in $a=N_f\alpha_1$. Physical
situations analogous to this exist in QCD.  For example a color singlet (Higgs)
scalar $\phi$ could decay to hadronic final states via the interaction ${\cal
L}_{\rm int} = \lambda g^2 \phi \tr\ G_{\mu\nu} G^{\mu\nu}$. The non-abelian
structure of QCD plays no role in the bubble chain sum of \bubble, so we have
simplified the computation by choosing an Abelian gauge theory, with two
different $U(1)$'s.

The total $\phi$ decay rate can be computed using the operator product
expansion, and the computation is very similar to the computation of $R(M^2)$.
The leading operator is $1$, and the first correction comes from dimension four
operators. One can also compute the Borel singularities using the methods of
ref.~\beneke. It is straightforward to show that the total decay rate has no
singularities in the Borel plane.

A more interesting computation is that of event shapes in $\phi$ decay. There
is no operator product expansion for the event shapes, so it is interesting to
see whether there exist renormalon singularities, and whether they are at
integer values of $u$. A convenient characterization of event shapes in
$\phi$ decay is the energy-energy correlation \ellis
\eqn\eedef{\eqalign{
&{\cal P} \left(\cos \chi\right) =\cr
& \sum_{ij} \int {d^3 \Gamma\over d E_i\
d E_j\ d \cos \theta_{ij}} \left({E_i\over M}\right) \left({E_j\over
M}\right) \delta \left(\cos \theta_{ij} - \cos \chi\right) dE_i d E_j d
\cos \theta_{ij}\cr
&\qquad\qquad+\delta \left(1 - \cos \chi \right) \sum_i \int
{d\Gamma\over dE_i} \left({E_i\over M}\right)^2 d E_i.
}}
In eq.~\eedef, the double sum is over all pairs $ij$ of particles in the final
state with $i\not=j$,  and $ij$ and $ji$ are both included in the sum. The
angle between the three-momentum vectors of particles $i$ and $j$ in the $\phi$
rest frame is denoted by $\theta_{ij}$, and $E_i$ represents the energy of
particle $i$. The Fox--Wolfram moments \ref\foxwolfram{G.C.~Fox and
S.~Wolfram, \prl{41} {1978} {1581} }\ $H_L$ are obtained
as integrals of the energy-energy correlations with Legendre polynomials,
\eqn\fox{
\int_{-1}^1 d\! \cos \chi\ {\cal P} \left(\cos \chi\right)\ P_L \left(\cos
\chi\right) = \Gamma \ H_L,
}
where $\Gamma$ is the total $\phi$ decay rate. Integrals of ${\cal P} (\cos
\chi)$ against any ``smooth'' weighting function $w (\cos \chi)$ are infrared
safe event shape variables.

The Feynman diagrams of \fig\shapefig{Feynman diagrams contributing to the
event shape distribution to lowest order in $\lambda$, $\alpha_1$ and
$\alpha_2$, and to all orders in $N_f\alpha_1$.}\ contribute to the $\phi$
decay distribution to lowest order in $\lambda, \alpha_1$ and
$\alpha_2$, and to all orders in $a=N_f\alpha_1$. We will restrict our analysis
to shape variables that can be obtained by integrating ${\cal P}$ against a
weighting function that vanishes at $\cos \chi = \pm 1$.  This simplifies the
calculation because the Feynman diagram in fig.~(2a) and the single sum
in eq.~\eedef\ don't contribute. The remaining Feynman graphs are easy to
compute, and one obtains
\eqn\pans{
{\cal P} \left(\cos \chi\right) = \int_0^{M^{2}} dq^2\ {d{\cal P} (\cos
\chi)\over dq^2},
}
where
\eqn\pqans{
{d{\cal P}\left(\cos \chi\right)\over dq^2} = {2M \lambda^2
\alpha_{2}\over N_f} {a^2\over \left|1 + \Pi (q^2) \right|^2} F\left(q^2,
\cos \chi\right).
}
Here $q^2$ is the invariant mass of the fermion pair,
\eqn\pidef{
\Pi(q^2) = -a\, b_0\, \ln \left(-q^2/\Lambda^2\right),\qquad
b_0 = 1/3\pi,
}
is the fermion bubble contribution to the gauge propagator, and
\eqn\fdef{\eqalign{
&F\left(q^2, \cos \chi\right) = {\hat q^2 \left(1-\hat q^2\right)^4
\left[ \left(1+\cos \chi \right)^2 + \hat q^4 \left(1 - \cos \chi
\right)^2 \right]\over \left[ \left(1 + \cos \chi \right) + \hat q^2
\left(1 - \cos \chi \right) \right]^5}\cr
&\quad + {\theta \left( \left(1 + \hat q^2 \right)^2 \left(1-\cos \chi \right)
- 8 \hat q^2 \right)\hat q^2 \left[ \left(1 - \cos \chi \right) \left(1
+ \hat q^4 \right) - 4 \hat q^2 \right]\over
\left( 1 - \cos \chi \right)^{7/2} \left[ \left(1 + \hat q^2
\right)^2 \left(1 - \cos \chi \right) - 8 \hat q^2 \right]^{1/2}},
}}
with $\hat q^2 = q^2/M^2$.  In eq.~\fdef\ terms
proportional to $\delta \left(1 + \cos \chi \right)$ and $\delta \left(1
- \cos \chi \right)$ are neglected. The quantity $\Lambda$ in eq.~\pidef\ is
proportional to the subtraction point $\mu$ used to define the finite
part of the vacuum polarization, $\Pi$, and is in general
subtraction scheme dependent.

We are interested in the Borel transform of ${\cal P} \left(\cos \chi
\right)$ with respect to the variable $a=N_f\alpha_1$. The dependence of
${\cal P} \left(\cos \chi \right)$ on $a$ is in the function
\eqn\gdef{\eqalign{
&g \left(a \right) = {a^2 \over \left|1 + \Pi (q^2)\right|^2} = {a^2\over
\left(1 - a b_0 \ln \left( q^2/\Lambda^2 \right) \right)^2 + b_0^2 a^2 \pi^2},
\cr
&= \sum_{m = 0}^\infty {\left(-1 \right)^m \pi^{2m}\over \left(2m + 1 \right)!}
\sum_{n=2m}^\infty \left(n + 1\right) n ... \left(n - 2m + 1\right)
b_0^n a^{n + 2} \left(\ln \left(q^2/\Lambda^2 \right) \right)^{n - 2m}.
}}
The Borel transform of $g\left(a \right)$ is
\eqn\gborel{\eqalign{
B[g](t) &= \sum_{m=0}^\infty {\left(-1 \right)^m \pi^{2m}\over (2m+1)!}
\sum_{n=0}^\infty { 1 \over n!} b_0^{n+2m}
t^{n+2m+1} \left(\ln \left(q^2/\Lambda^2 \right) \right)^n,\cr
&= \left(q^2/\Lambda^2 \right )^{b_0t} \sin \left(\pi b_0t \right)
,\cr
&= \left(q^2/\Lambda^2 \right )^{u} \sin \left(\pi u \right)  ,
}}
using $u=b_0 t$.

Consider an infrared safe shape variable $S_\eta$ that corresponds to a
weighting function $w(\cos \chi)$ that behaves like $w(\cos \chi) \sim
(1 + \cos \chi)^\eta$ near $\cos \chi = - 1$.  The singularities in the
Borel transform of this shape variable that arise from the integration region
near $\cos \chi = -1$ are determined by the value of  $F(q^2, \cos
\chi)$ for small $q^2$ and $\cos \chi$ near $-1$. In this region
\eqn\fapprox{
F\left(q^2, \cos \chi \right) \sim {\hat q^2 \left[ \left(1+\cos \chi
\right)^2 + 4 \hat q^4 \right]\over
\left[ \left(1+\cos \chi \right) + 2 \hat q^2 \right]^5}.
}
Using eqs.~\pans--\fapprox\ we find that the Borel transform of $S_\eta$
has infrared renormalons. For $\eta < 1$ the singularity closest
to the origin is a simple pole at
\eqn\irpole{
u=-\eta\qquad \Rightarrow\qquad t = - \eta/b_0.
}
For $\eta = 1$ there is potentially a simple pole at $u = -1$. However the
$\sin (\pi u)$ factor cancels the pole at $u=-1$, and there is no singularity
in
$B[S_1]$ at $u=-1$. The singularity closest to the origin is a simple pole
at $u=-2$.

An integral of the form
\eqn\hint{
 \int_0^{M^2} dq^2 h \left( q^2 \right) B\left[g\right]
 =  \sin \left( \pi u \right) \int_0^{M^2} dq^2
h\left( q^2 \right) \left(q^2/\Lambda^2\right)^u
}
determines the singularities in $B[S_{\eta}]$. If $h(q^2) \sim \left( q^2
\right)^k$ for small $q^2$, there is no singularity
 provided $k$ is an integer, and one obtains a simple pole singularity at
$u=-k-1$  if $k$ is not an integer. There is a simple pole singularity
at $u=-k-1$ for integer $k$ if $h(q^2) \sim \left( q^2 \right)^k \ln \left( q^2
\right)$ for small $q$. The weighting function $\left(1+\cos\chi
\right)^\eta$ produces an effective $\left( q^2 \right)^{\eta-1}$
behavior\foot{Although $h(q^2)$ is singular as $q^2\rightarrow 0$, it is
integrable so that $S_\eta$ is infrared safe.} for $h(q^2)$
at small $q^2$, and hence has a simple pole singularity at $u=-\eta$.
For $\eta=1$ (or more precisely, for $w\left( \cos\chi \right) = \cos^2
\chi -1$ which vanishes at $\cos\chi = \pm 1$), $h(q^2)$ in
eq.~\hint\ has the form $c_0 + c_1\ q^2 \ln q^2$ for small $q^2$,
where $c_0$ and $c_1$ are constants. The absence of a $\ln q^2$ term
(which is infrared safe) implies that there is no renormalon at $u=-1$.
The $q^2 \ln q^2$ term produces the pole at $u=-2$. In more
realistic examples, one expects that there will be a $\ln q^2$ term, and
hence a renormalon singularity at $u=-1$ as well.

The behavior of QCD hadronic event shape variables can be inferred from
the toy example using the method of Beneke \beneke. Our results imply
that Fox--Wolfram moments will have renormalons at integer values of
$u$. Event shape variables defined by weighting functions that are
polynomials in $\cos \chi $ (e.g. Fox-Wolfram moments) have Borel
transforms that are more singular than the Borel transform of the total
rate. More interesting are the shape variables $S_\eta$ ($0< \eta<1$)
which have a renormalon at $u=-\eta$. These variables have ambiguities
of order $(\Lambda_{QCD}^2/M^2)^\eta$ from the high-order sum of the
perturbation expansion.  Non-perturbative power suppressed corrections
are expected to be of the same order and cancel this ambiguity, leaving
a residual power suppressed correction.  The unusual power dependence of
the shape variables $S_\eta$ arises from
weighting functions $w(\cos \chi)$ that are continuous but not
differentiable at $\cos \chi = - 1$. Our results indicate that hadronic
shape variables which are infrared  safe but do not treat the region
where infrared divergences cancel between Feynman diagrams in a very
smooth fashion can have large power suppressed corrections.
  We hope to
present results for more realistic situations in a future publication.

\centerline{{\bf Acknowledgements}}
This work was supported in part by the Department of Energy under grant numbers
DOE-FG03-90ER40546 and DE-FG03-92-ER 40701. A.M. was also supported by
PYI award PHY-8958081.

\listrefs
\listfigs
\midinsert
\insertfig{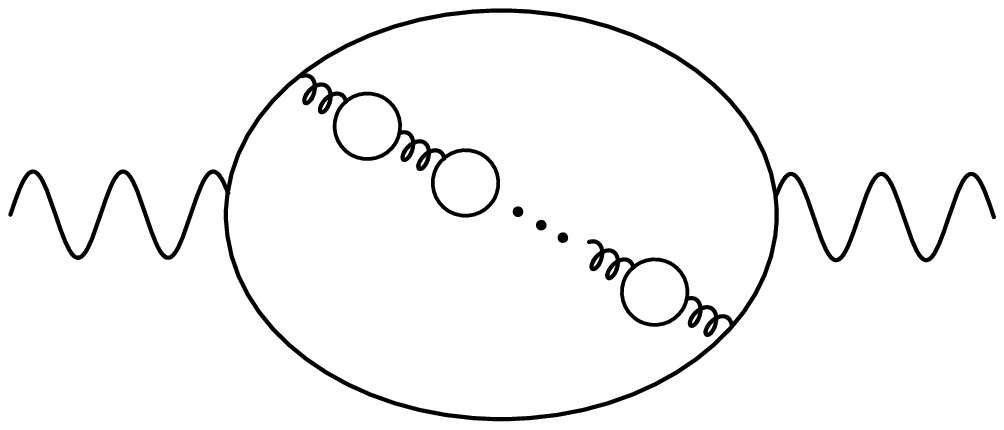}
\vskip0.75in\centerline{Figure 1}
\endinsert
\vskip1in
\midinsert
\insertfig{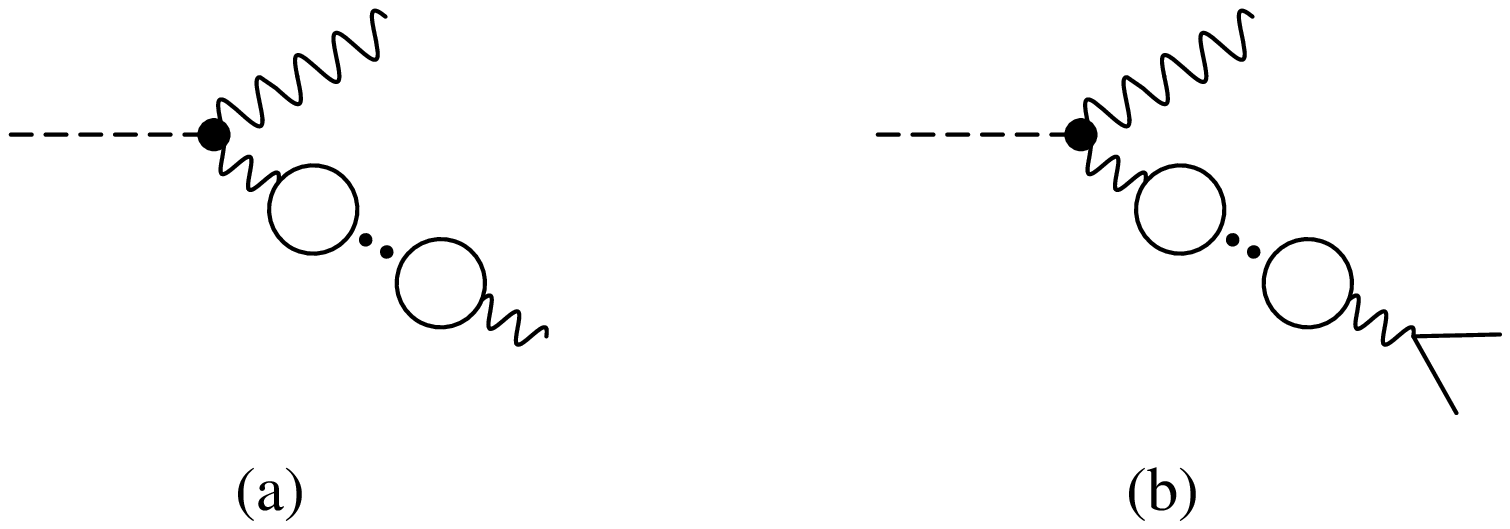}
\vskip1in\centerline{Figure 2}
\endinsert

\bye